# Effective t-J Hamiltonian for the Copper Oxides.

J.J. RODRIGUEZ-NUÑEZ[1,2] and HANS BECK

Université de Neuchâtel, Institut de Physique.
Rue A.L. Breguet 1, CH-2000 Neuchâtel.
**SUISSE.**

**Abstract**

Starting from the Emery model, which is assumed to describe the copper oxygen planes, and including direct oxygen hopping matrix elements, $t_{pp}$, we have been able to derive the effective $t - J$ Hamiltonian for the copper orbitals using the Linked Cluster Expansion Method up to fourth order in $t_{pd}$, the hybridization matrix element. The spin depend part of the effective Hamiltonian is composed of two contributions: the superexchange interaction and another one of RKKY-type. The effective parameters ($t$ and $J$) depend on doping, $\delta$. This Hamiltonian can be used to study the magnetic properties of the High-Tc materials vs. $\delta$.

[1]Permanent address: Departamento de Física. Grupo de Sólidos. Facultad Experimental de Ciencias. Universidad del Zulia. Apartado 526. Maracaibo. Estado Zulia. **VENEZUELA**
[2]Also a Visiting Scientist at IVIC. Centro de Física. Apartado 21827. Caracas 1020-A. **VENEZUELA.**

The study of High-Tc materials continue to attract attention both from the theoretical as from the experimental point of view.[1] There is general agreement that the basic structure of these materials is composed of the $CuO_2$ planes. There is evidence from both band - structure calculations and spectroscopy investigations that the generic properties of the $CuO_2$ planes can be described only in terms of the $Cu3d_{x^2-y^2}$ and the $Op_{x(y)}$ orbitals. These considerations lead us to the three - band Hubbard or Emery model. This is given by:[2]

$$H \equiv (\epsilon_d - \mu) \sum_{i\sigma} d^\dagger_{i\sigma} d_{i\sigma} + (\epsilon_p - \mu) \sum_{l\sigma} p^\dagger_{l\sigma} p_{l\sigma} + U_d \sum_i d^\dagger_{i\uparrow} d_{i\uparrow} d^\dagger_{i\downarrow} d_{i\downarrow}$$
$$+ t_{pp} \sum_{<l,l'>\sigma} (-1)^{N_{ll'}} p^\dagger_{l\sigma} p_{l'\sigma} + t_{pd} \sum_{<l,i>\sigma} (-1)^{M_{li}} \left( p^\dagger_{l\sigma} d_{i\sigma} + H.c. \right) , \quad (1)$$

where $t_{pd}$ is the hopping matrix element between the Cu and O orbitals, $U_d$ is the Coulomb repulsion of two holes in a Cu $d_{x^2-y^2}$ orbital, $\Delta \equiv \epsilon_p - \epsilon_d$, or charge transfer gap (CT), between the p and d orbitals, in the hole picture and $t_{pp}$ is the direct oxygen hopping matrix element. These are the most important parameters of the model. These parameters are usually taken from LDA calculations[3] to be : $\Delta \approx$ 3.5 eV, $t_{pd} \approx 1.35$ eV, $t_{pp} \approx 0.65$ eV, and $U_d \approx 9$eV. In Eq. (1), the indices **l** and **i** refer to the oxygen and copper sites, respectively. $\mu$ is the chemical potential. The phase factors $N_{ll'}$ and $M_{li}$ are either 0 or 1.[4]

The purpose of this letter is to give a derivation of the $t-J$[5] starting from the Emery model of Eq. (1). The $t-J$ has excitations which correspond to the low - energy excitations of the Emery model.[1]

To start our derivation, we Fourier analyze the oxygen orbitals. This gives:

$$H \equiv (\epsilon_d - \mu) \sum_{i\sigma} d^\dagger_{i\sigma} d_{i\sigma} + \sum_{k\nu\sigma} \epsilon_{k\nu} c^\dagger_{k\nu\sigma} c_{k\nu\sigma} + U_d \sum_i d^\dagger_{i\uparrow} d_{i\uparrow} d^\dagger_{i\downarrow} d_{i\downarrow} + H' , \quad (2)$$

where $H'$ is given by

$$H' \equiv \frac{t_{pd}}{\sqrt{N}} \sum_{ik\nu\sigma} \gamma_{k\nu} \left( e^{-i\mathbf{k}\cdot\mathbf{R}_i} d^\dagger_{i\sigma} c_{k\nu\sigma} + H.c. \right) \quad (3)$$

and

$$\epsilon_{k\nu} \equiv (\epsilon_p - \mu) + 4t_{pp}(-1)^\nu sin\frac{k_x}{2} sin\frac{k_y}{2} \;\; ; \;\; \gamma_{k\nu} \equiv \sqrt{2}(sin\frac{k_x}{2} + (-1)^\nu sin\frac{k_y}{2})^{1/2} \;\; , \quad (4)$$

In Eq. (3), $N$ is the number of copper sites. This model is the three band Anderson lattice model with the explicit oxygen band, $\epsilon_{\mathbf{k}\nu}$.[4] In contrast to the heavy - fermion systems, the oxygen bandwidth, $4t_{pp}$, has about the magnitude of $t_{pd}$ and the oxygen bands are lightly filled for $\delta \to 0$.

It can be shown that it is a good approximation to reduce the two oxygen bands to a single oxygen band.[6] This is done by means of a rotation given by $c_{\mathbf{k}\sigma} \equiv \gamma_{\mathbf{k}}^{-1} M_{\mathbf{k}} \phi_{\mathbf{k}\sigma}$, where $M_{\mathbf{k}}$ has the following form

$$M_{\mathbf{k}} = \begin{bmatrix} \gamma_{\mathbf{k}1} & \gamma_{\mathbf{k}2} \\ \gamma_{\mathbf{k}2} & -\gamma_{\mathbf{k}1} \end{bmatrix}$$

where

$$\gamma_{\mathbf{k}}^2 \equiv \gamma_{\mathbf{k}1}^2 + \gamma_{\mathbf{k}2}^2 = 4\left(sin^2(\frac{k_x}{2}) + sin^2(\frac{k_y}{2})\right). \tag{5}$$

Then, we end up with the following two band Hubbard Hamiltonian

$$H \equiv (\epsilon_d - \mu)\sum_{\mathbf{i}\sigma} d^\dagger_{\mathbf{i}\sigma} d_{\mathbf{i}\sigma} + \sum_{\mathbf{k}\sigma} \epsilon_{\mathbf{k}} \phi^\dagger_{\mathbf{k}\sigma}\phi_{\mathbf{k}\sigma} + U_d \sum_{\mathbf{i}} d^\dagger_{\mathbf{i}\uparrow}d_{\mathbf{i}\uparrow}d^\dagger_{\mathbf{i}\downarrow}d_{\mathbf{i}\downarrow} + H' \quad , \tag{6}$$

where $H'$ is given by

$$H' \equiv \frac{t_{pd}}{\sqrt{N}} \sum_{\mathbf{i}\mathbf{k}\sigma} \gamma_{\mathbf{k}} \left(e^{-i\mathbf{k}\cdot\mathbf{R}_{\mathbf{i}}} d^\dagger_{\mathbf{i}\sigma}\phi_{\mathbf{k}\sigma} + H.c.\right) \tag{7}$$

and

$$\epsilon_{\mathbf{k}} \equiv (\epsilon_p - \mu) + 4t_{pp}|sin\frac{k_x}{2}sin\frac{k_y}{2}| \quad \gamma_{\vec{k}} \equiv 2(sin^2\frac{k_x}{2} + sin^2\frac{k_y}{2})^{1/2} \quad , \tag{8}$$

where $|F|$ means absolute value of F. Let us mention that we are going to work with one oxygen band for simplicity. The case of two oxygen bands can be treated similarly, but the final expressions become more involved. We are trying to keep the algebra as simple as possible.

Next, we want to obtain an effective Hamiltonian for the coupling between the copper atoms. For this purpose we use the Linked Cluster Expansion Method.[7] This theorem tells us that the thermodynamic potential, $\Omega$, can be calculated from the following expression:

$$\Omega \equiv \Omega_o - \frac{1}{\beta}\sum_{n=1}^{\infty} U_n \quad , \tag{9}$$

where $\Omega_o$ is the unperturbed thermodynamic potential, $\beta$ is the inverse of the temperature and $U_n$ is given by

$$U_n \equiv \frac{(-1)^n}{n} \int_o^\infty d\tau_1 ... \int_o^\infty d\tau_n < T_\tau H'(\tau_1)...H'(\tau_n) >_c \quad , \tag{10}$$

and $T_\tau$ means time ordering.[7] The averages are performed with the oxygen orbitals and c means different conected diagrams.

Since,

$$< \phi_{\mathbf{k}\sigma}^{(2n+1)}(\tau) >_c \equiv < \phi_{\mathbf{k}\sigma}^{\dagger(2n+1)}(\tau) >_c \equiv 0 \quad , \tag{11}$$

only even terms of Eq. (10) contribute. Thus, in second order in $t_{pd}$, we have, with $H_2 \equiv U_2$,

$$H_2 \equiv -\frac{t_{pd}^2}{2\beta N} \sum_{\substack{\mathbf{i}\mathbf{k}\sigma \\ \mathbf{i}'\mathbf{k}'\sigma'}} \int_0^\beta d\tau_1 \int_0^\beta d\tau_2 \gamma_{\mathbf{k}} \gamma_{\mathbf{k}'} \left[ e^{-i(\mathbf{k}\cdot\mathbf{R}_{\mathbf{i}} + \mathbf{k}'\cdot\mathbf{R}'_{\mathbf{i}'})} \langle T_\tau \phi_{\mathbf{k}\sigma}^\dagger(\tau_1) \phi_{\mathbf{k}\sigma}^\dagger(\tau_2) \rangle d_{\mathbf{i}\sigma}(\tau_1) d_{\mathbf{i}'\sigma'}(\tau_2) \right.$$

$$+ e^{-i(\mathbf{k}\cdot\mathbf{R}_{\mathbf{i}} - \mathbf{k}'\cdot\mathbf{R}'_{\mathbf{i}'})} \langle T_\tau \phi_{\mathbf{k}\sigma}^\dagger(\tau_1) \phi_{\mathbf{k}\sigma}(\tau_2) \rangle d_{\mathbf{i}\sigma}(\tau_1) d_{\mathbf{i}'\sigma'}^\dagger(\tau_2)$$

$$+ e^{i(\mathbf{k}\cdot\mathbf{R}_{\mathbf{i}} - \mathbf{k}'\cdot\mathbf{R}'_{\mathbf{i}'})} \langle T_\tau \phi_{\mathbf{k}\sigma}(\tau_1) \phi_{\mathbf{k}\sigma}^\dagger(\tau_2) \rangle d_{\mathbf{i}\sigma}^\dagger(\tau_1) d_{\mathbf{i}'\sigma'}(\tau_2)$$

$$\left. + e^{i(\mathbf{k}\cdot\mathbf{R}_{\mathbf{i}} + \mathbf{k}'\cdot\mathbf{R}'_{\mathbf{i}'})} \langle T_\tau \phi_{\mathbf{k}\sigma}(\tau_1) \phi_{\mathbf{k}\sigma}(\tau_2) \rangle d_{\mathbf{i}\sigma}^\dagger(\tau_1) d_{\mathbf{i}'\sigma'}^\dagger(\tau_2) \right] \quad . \tag{12}$$

But,

$$\langle \phi^\dagger \phi^\dagger \rangle_c \equiv \langle \phi \phi \rangle_c \equiv 0 \quad , \tag{13}$$

$$\langle T_\tau \phi_{\mathbf{k}\sigma}^\dagger(\tau_1) \phi_{\mathbf{k}'\sigma'}(\tau_2) \rangle \equiv \delta_{\mathbf{k}\mathbf{k}'} \delta_{\sigma\sigma'} \langle T_\tau \phi_{\mathbf{k}\sigma}^\dagger(\tau_1) \phi_{\mathbf{k}\sigma'}(\tau_2) \rangle \quad , etc \tag{14}$$

Thus, Eq. (11) reduces to:

$$H_2 \equiv -\frac{t_{pd}^2}{2\beta N} \sum_{\substack{\mathbf{i}\mathbf{k}\sigma \\ \mathbf{i}'\mathbf{k}'\sigma'}} \int_0^\beta d\tau_1 \int_0^\beta d\tau_2 \gamma_{\mathbf{k}}^2 \left[ e^{-i\mathbf{k}\cdot\mathbf{R}_{\mathbf{i}\mathbf{i}'}} < T_\tau \phi_{\mathbf{k}\sigma}^\dagger(\tau_1) \phi_{\mathbf{k}\sigma}(\tau_2) >_c d_{\mathbf{i}\sigma}(\tau_1) d_{\mathbf{i}\sigma}^\dagger(\tau_2) \right.$$

$$\left. + e^{i\mathbf{k}\cdot\mathbf{R}_{\mathbf{i}\mathbf{i}'}} < T_\tau \phi_{\mathbf{k}\sigma}(\tau_1) \phi_{\mathbf{k}\sigma}^\dagger(\tau_2) >_c d_{\mathbf{i}\sigma}^\dagger(\tau_1) d_{\mathbf{i}\sigma}(\tau_2) \right] \quad . \tag{15}$$

Since,

$$G_c^o(\mathbf{k}, \tau) \equiv - < T_\tau \phi_{\mathbf{k}\sigma}(\tau_1) \phi_{\mathbf{k}\sigma}^\dagger(\tau_2) >_c \quad . \tag{16}$$

So, Eq. (11) is transformed to,

$$H_2 \equiv \sum_{\mathbf{i}\langle\mathbf{R}\rangle\sigma} t_{eff}(\mathbf{R}) \left( d_{\mathbf{i}\sigma}^\dagger d_{\mathbf{i}+\mathbf{R}\sigma} + H.C. \right) \quad , \tag{17}$$

where, the effective hopping matrix element, $t_{eff}(\mathbf{R})$ is given by

$$t_{eff}(\mathbf{R}) \equiv \frac{t_{pd}^2}{2N} \sum_{\mathbf{k}} \gamma_{\mathbf{k}}^2 \cos(\vec{k} \cdot \vec{R}) \left[ \frac{1}{(\epsilon_{\vec{k}} - \epsilon_d)} (n_F(\epsilon_d) - n_F(\epsilon_{\vec{k}})) \right.$$
$$\left. - \frac{1}{(\epsilon_{\vec{k}} - \epsilon_d - U_d)} (n_F(\epsilon_d + U_d) - n_F(\epsilon_{\vec{k}})) \right] , \quad (18)$$

with $n_F(\epsilon)$ being the Fermi - Dirac distribution function and $\langle \mathbf{R} \rangle$ means nearest neighbor atoms (n.n.).

Next, let us evaluate the fourth order term, $H_4 \equiv U_4$. It is given by

$$H_4 \equiv -\frac{1}{4\beta} \int_0^{\beta_1} \cdots \int_0^{\beta_4} < T_\tau H^{'}(\tau_1) \cdots H^{'}(\tau_4) >_c \quad . \quad (19)$$

After a straighforward calculation, we arrive to the following result:

$$H_4 \equiv \frac{t_{pd}^4}{4N^2} \sum_{\substack{\mathbf{i}\mathbf{k}\mathbf{p}\sigma \\ \mathbf{i}'\mathbf{k}'\mathbf{p}'\sigma'}} (\gamma_{\mathbf{k}} \gamma_{\mathbf{k}'})^2 \, \tilde{A}(\mathbf{k}, \mathbf{k}'; \beta) \left[ e^{-i(\mathbf{k} \cdot \mathbf{R}_{\mathbf{i}\mathbf{p}'} + \mathbf{k}' \cdot \mathbf{R}_{\mathbf{i}'\mathbf{p}})} d_{\mathbf{i}\sigma} d^\dagger_{\mathbf{p}\sigma} d_{\mathbf{i}'\sigma'} d^\dagger_{\mathbf{p}'\sigma} \right.$$
$$+ e^{-i(\mathbf{k} \cdot \mathbf{R}_{\mathbf{i}\mathbf{p}} + \mathbf{k}' \cdot \mathbf{R}_{\mathbf{i}'\mathbf{p}'})} d_{\mathbf{i}'\sigma'} d^\dagger_{\mathbf{p}\sigma} d_{\mathbf{i}\sigma} d^\dagger_{\mathbf{p}'\sigma'} + e^{-i(\mathbf{k} \cdot \mathbf{R}_{\mathbf{i}\mathbf{i}'} + \mathbf{k}' \cdot \mathbf{R}_{\mathbf{p}\mathbf{p}'})} d_{\mathbf{i}\sigma} d^\dagger_{\mathbf{p}'\sigma'} d_{\mathbf{p}\sigma'} d^\dagger_{\mathbf{i}'\sigma}$$
$$+ e^{-i(\mathbf{k} \cdot \mathbf{R}_{\mathbf{i}\mathbf{p}'} + \mathbf{k}' \cdot \vec{R}_{\mathbf{p}\mathbf{i}'})} d_{\mathbf{i}\sigma} d^\dagger_{\mathbf{i}'\sigma'} d_{\mathbf{p}\sigma'} d^\dagger_{\mathbf{p}'\sigma} + e^{-i(\mathbf{k} \cdot \mathbf{R}_{\mathbf{i}\mathbf{p}} + \mathbf{k}' \cdot \vec{R}_{\mathbf{p}'\mathbf{i}'})} d_{\mathbf{i}\sigma} d^\dagger_{\mathbf{i}'\sigma'} d_{\mathbf{p}'\sigma'} d^\dagger_{\mathbf{p}\sigma}$$
$$\left. + e^{-i(\mathbf{k} \cdot \mathbf{R}_{\mathbf{i}\mathbf{i}'} + \mathbf{k}' \cdot \mathbf{R}_{\mathbf{p}'\mathbf{p}})} d_{\mathbf{i}\sigma} d^\dagger_{\mathbf{p}\sigma'} d_{\mathbf{p}'\sigma'} d^\dagger_{\mathbf{i}'\sigma} + H.C. \right] , \quad (20)$$

where

$$\tilde{A}(\mathbf{k}, \mathbf{k}'; \beta) \equiv \frac{1}{\beta} \sum_{ik_n} G_c^o(\mathbf{k}, ik_n) G_c^o(\mathbf{k}', ik_n) g_d^2(ik_n) , \quad (21)$$

and

$$G_c^o(\mathbf{k}, ik_n) \equiv \frac{1}{(ik_n - \epsilon_{\mathbf{k}})} ; \quad (22)$$

$$g_d(ik_n) \equiv \frac{1}{2(ik_n - \epsilon_d)} + \frac{1}{2(ik_n - \epsilon_d - U_d)} , \quad (23)$$

where we have taken the atomic limit for the copper sites.[8]

By performing the Matsubara summation, we obtain

$$\tilde{A}(\mathbf{k}, \mathbf{k}'; \beta) \equiv \tilde{A}_1(\mathbf{k}, \mathbf{k}'; \beta) + \tilde{A}_2(\mathbf{k}, \mathbf{k}'; \beta) , \quad (24)$$

with

$$\tilde{A}_1(\mathbf{k},\mathbf{k}';\beta) \equiv \frac{n_F(\epsilon_d)}{2(\epsilon_{\mathbf{k}}-\epsilon_d)(\epsilon_{\mathbf{k}'}-\epsilon_d)} \left[\frac{1}{(\epsilon_{\mathbf{k}'}-\epsilon_d)} - \frac{1}{U_d}\right]$$
$$\frac{n_F(\epsilon_d+U_d)}{2(\epsilon_d+U_d-\epsilon_{\mathbf{k}})(\epsilon_d+U_d-\epsilon_{\mathbf{k}'})} \left[\frac{1}{(\epsilon_{\mathbf{k}}-\epsilon_d-U_d)} + \frac{1}{U_d}\right] \quad (25)$$

$$\tilde{A}_2(\mathbf{k},\mathbf{k}';\beta) \equiv \frac{n_F(\epsilon_{\mathbf{k}})}{2(\epsilon_{\mathbf{k}}-\epsilon_{\mathbf{k}'})} \left[\frac{1}{(\epsilon_{\mathbf{k}}-\epsilon_d)^2} + \frac{1}{(\epsilon_{\mathbf{k}}-\epsilon_d-U_d)^2} + \frac{2}{(\epsilon_{\mathbf{k}}-\epsilon_d)(\epsilon_{\mathbf{k}}-\epsilon_d-U_d)}\right], \quad (26)$$

In order to obtain a $\cos(\mathbf{k}-\mathbf{k}')\cdot\mathbf{R}$ dependence from Eq. (20), we need to make the following identifications

$$\mathbf{i}' \equiv \mathbf{p}' \; ; \; \mathbf{p} \equiv \mathbf{i} \quad \text{(first term)}.$$
$$\mathbf{i}' \equiv \mathbf{p} \; ; \; \mathbf{p}' \equiv \mathbf{i} \quad \text{(second term)}.$$
$$\mathbf{i}' \equiv \mathbf{p} \; ; \; \mathbf{p}' \equiv \mathbf{i} \quad \text{(third term, etc.)}$$

Then, we arrive to the following result

$$H_4 \equiv \frac{1}{2} \sum_{\substack{\mathbf{i}\langle\mathbf{R}\rangle \\ \sigma\sigma'}} J(\mathbf{R}) d_{\mathbf{i}\sigma} d^{\dagger}_{\mathbf{i}\sigma'} d_{\mathbf{i}+\mathbf{R}\sigma'} d^{\dagger}_{\mathbf{i}+\mathbf{R}\sigma} , \quad (27)$$

where

$$J(\mathbf{R}) \equiv \frac{3t_{pd}^4}{N^2} \sum_{\mathbf{k}\mathbf{k}'} \tilde{A}(\mathbf{k},\mathbf{k}';\beta) (\gamma_{\mathbf{k}}\gamma_{\mathbf{k}'})^2 \cos(\mathbf{k}-\mathbf{k}')\cdot\mathbf{R} . \quad (28)$$

On the other hand, the spin summation can be rewritten as

$$\sum_{\sigma\sigma'} d_{\mathbf{i}\sigma} d^{\dagger}_{\mathbf{i}\sigma'} d_{\mathbf{j}\sigma'} d^{\dagger}_{\mathbf{j}\sigma} \equiv \frac{1}{2} n_{\mathbf{i}} n_{\mathbf{j}} + 2\mathbf{S}_{\mathbf{i}}\cdot\mathbf{S}_{\mathbf{j}} . \quad (29)$$

So, $H_4$ can be expressed in a compact form as:

$$H_4 \equiv \sum_{\mathbf{i}\langle\mathbf{R}\rangle} J(\mathbf{R}) \left(\mathbf{S}_{\mathbf{i}}\cdot\mathbf{S}_{\mathbf{i}+\mathbf{R}} + \frac{1}{4} n_{\mathbf{i}} n_{\mathbf{i}+\mathbf{R}}\right) . \quad (30)$$

From Eq. (30), we can conclude that the fourth order contribution to the effective Hamiltonian has the form of spin dependent part $J(\mathbf{R})$, as in the $t-J$ model.

This $J(\mathbf{R})$ term has two contributions (See Eqs. (24), (25) and (26)). If we make the following approximation

$$\epsilon_{\vec{k}} - \epsilon_d \approx \Delta \ , \tag{31}$$

we see that $\tilde{A}_m(\vec{k}, \vec{k}'; \beta), \ m = 1, 2$ is given by

$$\tilde{A}_1(\mathbf{k}, \mathbf{k}'; \beta) \approx D_1 \ , \tag{32}$$

$$\tilde{A}_2(\mathbf{k}, \mathbf{k}'; \beta) \approx D_2 \frac{1}{(\epsilon_{\mathbf{k}} - \epsilon_{\mathbf{k}'})} \ , \tag{33}$$

where $D_1$ and $D_2$ are constants, with $D_1 > 0$.

$J(\mathbf{R})$ has two contributions: 1) a superexchange type of interaction due to $\tilde{A}_1(\mathbf{k}, \mathbf{k}'; \beta)$;[9] and 2) a RKKY-type of interaction [6] ($\tilde{A}_2(\mathbf{k}, \mathbf{k}'; \beta)$) which is strongly dependent upon doping rate, $\delta$ and has an oscillatory behavior at very long distance between the copper atoms. It is this doping dependency of this $J(\mathbf{R})$ which makes it interesting. It is possible to study the magnetic properties as function of doping, as it was done in ref. [6]. In particular, the $J_{RKKY}$ interaction might lead to a frustation of the long - range AF ordering with doping (small $\delta$). In addition to the $J(\mathbf{R})$ term, we find a local repulsion given by the $U_d$ term.

At this point, we have the effective Hamiltonian, $H_{eff}$, on the copper sites, given by

$$H_{eff} \equiv (\epsilon_d - \mu) \sum_{\mathbf{i}\sigma} d^{\dagger}_{\mathbf{i}\sigma} d_{\mathbf{i}\sigma} + H_2 + H_4 + U_d \sum_{\mathbf{i}} d^{\dagger}_{\mathbf{i}\uparrow} d_{\mathbf{i}\uparrow} d^{\dagger}_{\mathbf{i}\downarrow} d_{\mathbf{i}\downarrow} \ . \tag{34}$$

In the large $U_d$ limit, we can enforce no double occupancy at half filling, by using perturbation theory on virtual states.[10] The result is

$$\tilde{H}_{eff} \equiv (\epsilon_d - \mu) \sum_{\mathbf{i}\sigma} \tilde{d}^{\dagger}_{\mathbf{i}\sigma} \tilde{d}_{\mathbf{i}\sigma} + t \sum_{\mathbf{i}\langle\mathbf{R}\rangle\sigma} \tilde{d}^{\dagger}_{\mathbf{i}+\mathbf{R}\sigma} \tilde{d}_{\mathbf{i}\sigma}$$

$$+ (J_1 + J_2) \sum_{\mathbf{i}\langle\mathbf{R}\rangle} \mathbf{S}_{\mathbf{i}} \cdot \mathbf{S}_{\mathbf{i}+\mathbf{R}} + (J_1 - J_2) \sum_{\mathbf{i}\langle\mathbf{R}\rangle} n_{\mathbf{i}+\mathbf{R}} n_{\mathbf{i}} + \text{(three site terms)} \ , \tag{35}$$

where

$$t \equiv t_{eff}(\mathbf{R}) \ ; \ J_1 \equiv J(\mathbf{R}) \ ; \ J_2 \equiv \frac{4t^2}{U_d} \ , \tag{36}$$

$$\tilde{d}_{\mathbf{i}\sigma} \equiv d_{\mathbf{i}\sigma}(1 - n_{\mathbf{i}-\sigma}) \ . \tag{37}$$

In conclusion, we have been able to derive the $t - J$ Hamiltonian starting from the three band Hubbard or Emery model. Our effective parameter $J_1$ contains the superexchange interaction of the $t-J$ derived by Zhang and Rice,[5] in the limite of $t_{pp} \to 0$. However, $t_{pp}$ is a key element in the description of the physics of High-Tc superconductors and it should not be taken equal to zero. The effective parameters depend on doping by means of the Fermi level. The spin dependent part of the effective Hamiltonian is composed of two contributions, one mainly antiferromagnetic given by $\tilde{A}_1$, and a second one given by $\tilde{A}_2$, which is of the RKKY - type. To our knowledge, it is the first time that the latest contribution has been derived simultaneously with the superexchange contribution. There is a calculation by Eskes and Jefferson[11] where they have included $t_{pp}$ and then they use fifth order perturbation theory. However, they do not obtain the RKKY - type interaction. The presence of these two contributions are interesting for studying the evolution of the magnetic properties upon doping.

## Acknowlegments

The authors would like to thank **CONICIT** - Venezuela for finantial support throught the project No. F-139 and the Swiss National Foundation, project No. 4030-032788. It is a pleasure to thank Profs. A.M. Oleś, S. Barnes and Toni Schneider and Drs. R. Frezárd and Sudha Gopalan for interesting discussions on different parts of this research. Partial support from **CONDES** - Venezuela is also appreciated. We would like to thank María Dolores García-González for substantially improving the English of the original manuscript.

## References.